\documentclass[times,10pt]{IEEEtran11}
\twocolumn

\include{epsf}

\newenvironment{ctable*}
{\begin{table*}[htpb]\begin{center}}{\end{center}\end{table*}}

\newtheorem{theorem}{Theorem}[section]
\newtheorem{proposition}[theorem]{Proposition}
\newtheorem{corollary}[theorem]{Corollary}

\newtheorem{definitionx}[theorem]{Definition}

\newenvironment{definition}{\begin{definitionx}\rm}{\end{definitionx}}
\newenvironment{example}{\begin{examplex}\rm}{\end{examplex}}
\newtheorem{examplex}[theorem]{Example}

\newcommand{\prf}{\begin{proof}}
\newcommand{\eprf}{\end{proof}}
\newcommand{\Gmin}{G_{{\rm min}}}
\newcommand{\Emin}{E_{{\rm min}}}
\newcommand{\Loc}{\mbox{{\it Loc}}}
\newcommand{\NonNbrNodes}{\ensuremath{\mathit{NonNbrs}}}
\newcommand{\flipAllStatesDownChain}{{\it Flip}}
\newcommand{\flip}{{\it Flip}}

\newcommand{\Increase}{\mbox{{\it Increase}}}
\newcommand{\leave}{\mbox{{\it leave}}}
\newcommand{\move}{\mbox{{\it move}}}
\newcommand{\join}{\mbox{{\it join}}}
\newcommand{\union}{\ensuremath{\mathit{\cup}}}

\newcommand{\commentout}[1]{}

\include{epsf}

\begin{document}

\commentout{
\begin{figure*}[ht]
\begin{tabular}{ll}
Title:        & {\Large Minimum-Energy Mobile Wireless Networks Revisited} \\
              & \\
Authors:      & Li Li and Joseph Y. Halpern \\
Affiliations: & Dept. of Computer Science, \\
              & Cornell University \\
              & \\
Corresponding author: & \\
              & Li Li \\
              & 4161 Upson Hall \\
              & Dept. of Computer Science \\
              & Cornell University \\
              & Ithaca, NY, 14853-7501 \\
              & USA             \\ 
              & Tel: 607-2536487 \\
              & Fax: 607-2554485 \\
              & Email: lili@cs.cornell.edu 
\end{tabular}
\end{figure*}

\setcounter{page}{0}
}

\title{\Large\bf  
Minimum-Energy Mobile Wireless Networks Revisited \\
}


\author{
\begin{tabular}{cc}
Li Li & Joseph Y. Halpern \\
{ Dept. of Computer Science} & { Dept. of Computer Science} \\
{ Cornell University} & {Cornell University} \\
{ Ithaca NY 14853} & {Ithaca NY 14853}\\
{\tt   lili@cs.cornell.edu} & {\tt  halpern@cs.cornell.edu} 
\end{tabular}
}

\date{}

\maketitle
\thispagestyle{empty}
\pagestyle{empty}

\title{\Large\bf  Minimum Energy Mobile Wireless Networks Revisited \\
}
\begin{abstract}
We propose a protocol that,
given a communication network, computes a subnetwork such that, for
every pair $(u,v)$ of nodes connected in the original network, there is a 
a {\em minimum-energy path\/} between $u$ and $v$ in the subnetwork
(where a minimum-energy path is one that
allows messages to be transmitted with a minimum use of energy).
The network computed by our protocol is in general a
subnetwork of the one computed by the protocol given in
\cite{Rodoplu99}.  Moreover, our protocol is computationally simpler.
We demonstrate the performance improvements 
obtained by using  the subnetwork computed by our
protocol through simulation.        
\end{abstract}


\section{Introduction}
\label{sec-introduction}

Multi-hop wireless networks, especially sensor
networks, are expected to be deployed in a wide variety of civil and military
applications. 
Minimizing energy consumption
has been a major design goal for wireless networks.
As pointed out by Heinzelman et. al \cite{Chandrakasan99},
network protocols that minimizes energy consumption are key
to low-power wireless sensor networks. 

We can characterize a communication network using  a graph $G'$ where 
the nodes in $G'$ represent the nodes in the network, and two nodes
$u$ and $v$ are joined by an edge if it is possible for $u$ to transmit
a message to $v$ if $u$ transmits at maximum power.  
Transmitting at maximum power requires a great deal of energy.
To minimize energy usage, we would like a subgraph $G$ of $G'$ such that
(1) $G$ consists of all the nodes in $G'$ but has fewer edges, (2) if
$u$ and $v$ are connected in $G'$, they are still connected in $G$, and 
(3) a node $u$ can transmit to all its neighbors in $G$ using less
power than is required to transmit to all its neighbors in $G'$.
Indeed, what we would really like is a subnetwork $G$ of $G'$ with these
properties where the power for a node to transmit to its neighbors in
$G'$ is minimal.  
Rodoplu and Meng \cite{Rodoplu99} provide  a 
protocol
that, given a communication network, computes a subnetwork that is
energy-efficient in this sense.
We call their protocol MECN
(for {\em minimum-energy communication network\/}).

The key property of the subnetwork constructed by MECN is what we call
the {\em minimum-energy property}.  Given $G'$, it guarantees that
between every pair $(u,v)$ of nodes that are connected in $G'$, the
subgraph $G$ has a {\em minimum-energy path\/} between $u$ and $v$,
one that allows messages to be transmitted with a minimum use of energy
among all the paths between $u$ and $v$ in $G'$.
In this paper, we first identify 
conditions that are necessary and sufficient for 
a graph to have this minimum-energy property.
We use this characterization to construct a protocol
called SMECN (for {\em small\/} minimum-energy communication
network). 
The subnetwork constructed by SMECN is provably smaller than that
constructed by MECN if broadcasts at a given power setting are able to
reach all nodes in a circular region around the broadcaster.  We
conjecture that this property will hold in practice even without this
assumption.   
Our simulations show that
by being able to use a smaller network, 
SMECN has lower link maintenance costs than MECN
and can achieve a significant
saving in energy usage.
SMECN is also computationally 
simpler than MECN.

The rest of the paper is organized as follows. Section~\ref{sec-model}
gives the network model
(which is essentially the same as that used in \cite{Rodoplu99}).
Section~\ref{sec-char} identifies 
a condition necessary and sufficient for achieving the minimum-energy
property.
This characterization is used in Section~\ref{sec-SMECN} to construct
the SMECN protocol and prove that it constructs a network smaller than
MECN if the broadcast region is circular.
In Section~\ref{sec-sim}, we give the results of simulations showing the
energy savings obtained by using the network constructed by SMECN.
Section~\ref{sec-conclusion} concludes our paper.

\section{The Model}
\label{sec-model}
We use essentially the same model as 
Rodoplu and Meng \cite{Rodoplu99}. 
We assume that a set $V$ of nodes is deployed in a two-dimensional
area, where no two nodes are in the same physical location. 
Each node has a GPS receiver on board, so knows it own location, to
within at least 5 meters of accuracy.  It does not necessarily know the
location of other nodes.  Moreover, the location of nodes will in
general change over time.

\commentout{
Rodoplu and Meng
implicitly assume that a node can transmit directly to any other node.  In
practice, a node's signal can only propagate a certain distance. We
assume a maximum transmission radius $d_{max}$ for all nodes. 
Of cause, our results hold for $d_{max} = \infty$.

We define
the {\em reference network\/} to be $G'=(V,E')$, where $E'= \{ (u,v) |$,
$d(u,v)$ $\leq$ $d_{max}$ $\}$.  Thus, $G'$ characterizes which nodes
are, in principle, able to communicate with each other.  We assume that
$G'$ is connected, so that there is a potential communication path
between any pair of nodes in $V$.}

A transmission between node $u$ and $v$ takes power $p(u,v)$ $=$ $t
{d(u,v)}^n$ 
for some appropriate constant $t$, where
$n\ge 2$ is the path-loss exponent of outdoor radio
propagation models \cite{Rap96},  and $d(u,v)$ is the distance between
$u$ and $v$. A reception at the receiver takes power
$c$. Computational power consumption is ignored. 

Suppose there is some maximum power $p_{max}$ at which the nodes can transmit.
Thus, there is a graph $G' = (V,E')$ where $(u,v) \in E'$ if it is possible
for $u$ to transmit to $v$ if it transmits at maximum power.  
Clearly,
if $(u,v) \in E$, then $td(u,v)^n \le p_{max}$.  However, we do not assume
that a node $u$ can transmit to all  nodes $v$ such that $t d(u,v)^n \le
p_{max}$.  For one thing, there may be obstacles between $u$ and $v$
that prevent transmission.
Even without obstacles, if a unit transmits using a directional transmit
antenna, then only nodes in the region covered by the antenna (typically
a cone-like region) will receive the message.
Rodoplu and Meng \cite{Rodoplu99} implicitly assume that every
node can transmit to every other node.  Here we take a first step in
exploring what happens if this is not the case.  However, we do assume
that the graph $G'$ is connected, so that there is a potential
communication path between any pair of nodes in $V$.

Because the power required to transmit between a pair of nodes increases
as the $n$th power of the distance between them, for some $n \ge 2$, it
may require less power to relay information than to transmit directly
between two nodes.
As usual, a {\em path\/} $r = (u_0, \ldots, u_k)$ in a graph $G = (V,E)$ is
defined to be an ordered list of nodes such that $(u_i, u_{i+1}) \in E$.
The {\em length\/} of $r= (u_0, \ldots, u_k)$, denoted $|r|$, is $k$.
The total power consumption of a
path $r = (u_0,u_2,\cdots,u_k)$ in $G'$
is the sum of the transmission and reception power consumed,
i.e., $$C(r) = \sum_{i=1}^{k-1} (p(u_i,u_{i+1})+c).$$    
A path $r = (u_0,\ldots, u_k)$ is a {\em minimum-energy path\/} from
$u_0$ to $u_k$ if $C(r) \le C(r')$ for all paths $r'$ in $G'$ from $u_0$
to $u_k$.  
For simplicity, we assume that $c > 0$.  (Our results hold even without
this assumption, but it makes the proofs a little easier.)

A subgraph $G = (V,E)$ of $G'$ has the {\em minimum-energy
property\/} if, for all $(u,v) \in V$, there is a path $r$ in $G$ that
is a minimum-energy path in $G'$ from $u$ to $v$.

\section{A Characterization of Minimum-Energy Communication Networks}
\label{sec-char}
Our goal is to find a minimal subgraph $G$ of $G'$ that 
has the minimum-energy property.
Note that a graph $G$ with the minimum-energy property
must be 
strongly
connected since, by definition, it
contains a path between any pair of nodes.  Given such a graph, the 
nodes can communicate using the links in $G$. 

For this to
be useful in practice, it must be possible for each of the nodes in the
network to construct $G$ (or, at least, the relevant portion of $G$ from their
point of view) in a distributed way.  In this section, we provide a
condition that is necessary and sufficient for a subgraph of $G'$ to be minimal
with respect to the minimum-energy property.  In the next section, we
use this characterization to 
provide an efficient algorithm for constructing a graph $G$ with the
minimum-energy property that, while not 
necessarily
minimal, still has relatively
few edges.



Clearly if 
a subgraph $G= (V,E)$ of $G'$ has the minimum-energy property,
an edge $(u,v) \in E$ is 
{\em redundant\/}
if there is a path $r$ from
$u$ to $v$ in $G$ such that 
$|r| > 1$ and $C(r) \le C(u,v)$.  
Let $\Gmin = (V, \Emin)$ be the subgraph of $G'$ such that $(u,v) \in
\Emin$ iff there is no path $r$ from $u$ to $v$ in $G'$ such that $|r| >
1$ and $C(r) \le C(u,v)$.
As the next result shows, $\Gmin$ is the smallest subgraph of $G'$ with the
minimum-energy property.


\begin{theorem} 
\label{the-ifonlyif}
A subgraph $G$ of
$G'$ has the minimum-energy property iff it contains $\Gmin$ as a
subgraph. Thus, $\Gmin$ is the smallest subgraph of $G'$ with 
the minimum-energy property.
\end{theorem}
\prf
We first show that $\Gmin$ has the minimum-energy property.
Suppose, by way of contradiction,
that there are nodes $u, v \in V$ and a 
path $r$ in $G'$ from $u$ to $v$ such that $C(r) <
C(r')$ for any path $r'$ from $u$ to $v$ in $\Gmin$.  Suppose that $r =
(u_0, \ldots, u_k)$, where $u=u_0$ and $v = u_k$.  
Without loss of generality, we can assume that $r$ is the longest
minimal-energy path from $u$ to $v$.  Note that $r$ has no repeated
nodes 
for any cycle 
can be removed to give a path that
requires strictly less power.
Since $\Gmin$ has no redundant edges, for all $i = 0, \ldots, k-1$, 
it follows that
$(u_i, u_{i+1}) \in \Emin$.
For otherwise, 
there is a path $r_i$ in $G'$ from $u_i$ to
$u_{i+1}$ such that $|r_i| > 1$ and $C(r_i) \le C(u_i,u_{i+1})$.  But
then it is
immediate that there is a path $r^*$ in 
$G'$ such that $C(r^*) \le C(r)$ and $r^*$ is longer than $r$, 
contradicting the choice of $r$.  

To see that $\Gmin$ is a subgraph of every subgraph of $G'$ with the
minimum-energy property, suppose that there is some subgraph $G$ of $G'$
with the minimum-energy property that does not contain the edge $(u,v)
\in \Emin$.  Thus, there is a minimum-energy path
$r$ from $u$ to $v$ in $G$.  It must be the case
that $C(r) \le C(u,v)$.  Since $(u,v)$ is not an edge in $G$, we must
have $|r|>1$. 
But then $(u,v) \notin \Emin$, a contradiction.
\eprf


This result shows that in order to find a subgraph of $G$ with the
minimum-energy property, it suffices to ensure that it contains $\Gmin$
as a subgraph.

\section{A 
Power-Efficient
Protocol for Finding a Minimum-Energy Communication Network}
\label{sec-SMECN}
Checking if an edge $(u,v)$ is in $\Emin$ may require
checking nodes that are located far from $u$.  
This may require a great deal of communication, possibly to distant
nodes, and thus require a great deal of power.
Since power-efficiency is an important consideration in practice, we
consider here an algorithm for constructing a communication network
that contains $\Gmin$ and can be constructed in a power-efficient manner
rather than trying to construct $\Gmin$ itself.  

Say that an edge $(u,v) \in E'$ is {\em $k$-redundant\/}
if there is a path $r$ in $G'$ such that 
$|r| = k$ 
and $C(r) \le C(u,v)$.  Notice that $(u,v) \in \Emin$ iff it is not
$k$-redundant for all $k > 1$.  
Let $E_2$ consist of all and only edges in $E'$ that are not 2-redundant.
In our algorithm, we construct a graph $G = (V,E)$ where $E \supseteq
E_2$; in fact, under appropriate assumptions, $E = E_2$.
Clearly $E_2 \supseteq \Emin$, so
$G$ has the minimum-energy property.  

There is a trivial algorithm for constructing $E_2$.  Each node $u$ starts
the process by broadcasting a {\em neighbor discovery 
message\/} (NDM) at maximum power $p_{max}$, stating its own position. 
If a 
node $v$ receives this message, it responds to $u$ with a message
stating its location.  Let $M(u)$ be the set of nodes 
that respond to $u$ and let $N_2(u)$ denote $u$'s neighbors in $E_2$.
Clearly $N_2(u) \subseteq M(u)$.  Moreover, it is easy to check that 
$N_2(u)$ consists of all those nodes  $v \in M(u)$ other than $u$ such 
that there is no $w \in M(u)$ such that $C(u,w,v) \le C(u,v)$.  Since $u$
has the location of all nodes in $M(u)$, $N_2(u)$ is easy to compute.

The problem with this algorithm is in the first step, which involves a
broadcast using maximum power.  While this expenditure of
power may be necessary if there are relatively few nodes, so that power
close to $p_{max}$ will be required to transmit to some of 
$u$'s neighbors in $E_2$, it is
unnecessary in denser networks.  In this case, it may require much less
than $p_{max}$ to find $u$'s neighbors in $E_2$.
We now present a
more power-efficient algorithm for finding these neighbors, based on
ideas due to Rodoplu and Meng \cite{Rodoplu99}.
For this algorithm, we assume that if a node $u$ transmits with power
$p$, it knows the region $F(u,p)$ around $u$ which can be reached with
power $p$.  If there are no obstacles
and the antenna is omnidirectional,
then this region is just a circle
of radius $d_p$ such that $td_p^n = p$.  We are implicitly assuming that
even if there are obstacles 
or the antenna is not omni-directional,
a node $u$ knows the terrain 
and the antenna characteristics 
well enough to
compute $F(u,p)$.  If there are no obstacles, we show that 
$E_2$ is 
a subgraph of what
Rodoplu and Meng call the {\em enclosure graph}.
Our algorithm is a variant of their algorithm for constructing the
enclosure graph.

Before presenting the algorithm, it is useful to define a few terms.

%


\begin{definition} 
Given a node $v$, let $\Loc(v)$ denote the physical location of $v$.
The {\em relay region\/} of the transmit-relay node pair $(u,v)$ is the
physical region $R_{u \rightarrow v}$ such that relaying through $v$
to any point in $R_{u \rightarrow v}$ takes less power than direct
transmission.  Formally,
$$R_{u \rightarrow v} = \{(x,y): C(u,v,(x,y)) \le C(u,(x,y))\},$$
where we abuse notation and take $C(u,(x,y))$ to be 
the cost of transmitting a message from $u$ to a virtual node whose
location is $(x,y)$.  That is, if there were a node $v'$ such that
$\Loc(v') = (x,y)$, then $C(u,(x,y)) = C(u,v')$; similarly,
$C(u,v,(x,y)) = C(u,v,v')$.
Note that, if a node $v$ is
in the relay region $R_{u \rightarrow w}$, then 
the edge $(u,v)$ is 2-redundant.
Moreover, since $c > 0$, $R_{u \rightarrow u} = \emptyset$.

\commentout{
\footnote{The notion of relay region was introduced by Rodoplu and Meng
\cite{Rodoplu99}.  Our definition differs from theirs in 
one minor respect:  
we require only that
$C(u,v,(x,y)) \le C(u,(x,y))$.  Rodoplu and Meng use $<$ rather than $\le$.}
}
\end{definition}

Given a region $F$, let $$N_F = \{v \in V: \Loc(v) \in
F\};$$ if $F$ contains $u$, 
let 
\begin{equation}
\label{eq-RF}
R_F(u) = \bigcap_{w \in N_F}
(F(u,p_{max}) - R_{u\rightarrow w}).
\end{equation}

The following proposition gives a useful characterization of $N_2(u)$.
\begin{proposition}\label{N2char}
Suppose that $F$ is a region containing the node $u$.
If $F \supseteq R_F(u)$, then $N_{R_F(u)} \supseteq N_2(u)$.  
Moreover, if $F$ is a circular region with center $u$ and $F \supseteq
R_F(u)$, then $N_{R_F(u)} = N_2(u)$.
\end{proposition}
\begin{proof}
Suppose that $F$ $\supseteq$ $R_F(u)$.  We show that $N_{R_F(u)}$
$\supseteq$ $N_2(u)$.  Suppose that $v$ $\in$ $N_2(u)$.  Then clearly 
$\Loc(v)$ $\notin$ $\cup_{w \in V\,} R_{u \rightarrow w}$ 
and $\Loc(v)$ $\in$ $F(u,p_{max})$.
Thus, $\Loc(v)$ $\in$ $R_F(u)$, so $v$ $\in$ $N_{R_F(u)}$.

Now suppose that $F$ is a circular region with center $u$ and
$F \supseteq R_F(u)$.  The preceding paragraph
shows that $N_{R_F(u)} \supseteq N_2(u)$.
We now show that 
$N_{R_F(u)} \subseteq N_2(u)$.  Suppose that $v \in N_{R_F(u)}$.  If $v \notin
N_2(u)$, then there exists some $w$ such that $C(u,w,v) \le C(u,v)$.
Since transmission costs increase with distance, it must be the case
that $d(u,w) \le d(u,v)$.  Since $v \in N_{R_F(u)} \subseteq N_F$ and 
$F$ is a circular region with center
$u$, it follows that $w \in N_F$.    Since $C(u,w,v) \le C(u,v)$, it
follows that $\Loc(v) \in R_{u \rightarrow w}$.  Thus, $v \notin
R_F(u)$, contradicting our original assumption.  Thus, $v \in N_2(u)$.
\end{proof}

The algorithm for node $u$ constructs a set $F$ such that $F \supseteq
R_F(u)$, and tries to do so in a power-efficient fashion. 
By
Proposition~\ref{N2char}, the fact that $F \supseteq R_F(u)$ ensures
that $N_{R_F(u)} \supseteq N_2(u)$.  Thus, the nodes in $N_{R_F(u)}$
other than $u$ itself are taken to be $u$'s neighbors.  By 
Theorem~\ref{the-ifonlyif},  the resulting graph has the minimum-energy
property.  

Essentially, the algorithm for node $u$ starts by broadcasting an NDM
with some initial power $p_0$, getting responses from all nodes in
$F(u,p_0)$, and checking if $F(u,p_0) \supseteq R_{F(u,p_0)}(u)$.
If not, it transmits with more power.  It continues increasing the power
$p$ until $F(u,p) \supseteq R_{F(u,p)}(u)$.  It is easy to see that
$F(u,p_{max}) \supseteq R_{F(u,p_{max})}(u)$, so that as long as the
power increases to $p_{max}$ eventually, then this process
is guaranteed to terminate.  
In this paper, we do not investigate how to choose the initial power
$p_0$, nor do we investigate how to increase the power at each step.
We simply assume some function $\Increase$ such that $\Increase^k(p_0)
= p_{max}$ for sufficiently large $k$.  An obvious choice is to
take $\Increase(p) = 2p$.  If the initial choice of $p_0$ is less than
the total power actually needed, then it is easy to see that this
guarantees that  
the total amount of transmission power used by $u$ will be within a
factor of 2 of optimal.
\footnote{Note that, in practice, a node may control a number of
directional transmit antennae.  Our algorithm implicitly assumes that
they all transmit at the same power.  This was done for ease of
exposition.  It would be easy to modify the algorithm to allow each
antenna to transmit using different power.  All that is required is that
after sufficiently many iterations, all antennae transmit at
maximum power.}

Thus, the protocol run by node $u$ is simply
\begin{tabbing}
\ \ \ \ \ \ \=$p = p_0$;\\
\>while $F(u,p) \not\supseteq R_{F(u,p)}(u)$ do $\Increase(p)$;\\
\>$N(u) = N_{R_{F(u,p)}}$
\end{tabbing}
A more careful implementation of this algorithm is given 
in Figure \ref{fig-SMECN}.  
Note that we also compute the minimum power
$p(u)$
required to reach all the nodes in $N(u)$.
In the algorithm,  $A$ is the set of all the nodes that $u$ has found so
far in the search and $M$ consists of the new nodes found in the current
iteration. 
In the the computation of $\eta$ in the second-last line of the
algorithm, we take $\cap_{v \in M} ( F(u,p_{max}) - R_{u\rightarrow v})$
to be $F(u,p_{max})$ if $M = \emptyset$.  For future reference, we note
that it is easy to show that, after each iteration of the while loop, we
have that $\eta = \cap_{v \in A} ( F(u,p_{max}) - R_{u\rightarrow v})$.

\begin{figure}[htb]
\begin{tabbing}
\underline{Algorithm SMECN}   \\
$p = p_0$;\\
$A = \emptyset$; \\
$\NonNbrNodes = \emptyset;$ \\
$\eta = F(u,p_{max})$; \\
while \= $F(u,p) \not \supseteq \eta$ do\\
\>    $p = \Increase(p)$;\\
\>    Broadcast NDM with power $p$ and gather responses;\\
\>     $M = \{v | \Loc(v) \in F(u,p), v \not \in A, v \not= u\}$; \\
\>     $A = A \bigcup M$; \\
\>     for \=each $v \in M$ do \\
\> \>        for \=each $w \in A$ do \\
\> \> \>             if \= $\Loc(v) \in R_{u\rightarrow w}$ then \\
\> \> \> \>                  $\NonNbrNodes = \NonNbrNodes \bigcup \{v\}$; \\
\> \> \>             else if $\Loc(w) \in R_{u\rightarrow v}$ then \\
\> \> \> \>                  $\NonNbrNodes = \NonNbrNodes \bigcup \{w\}$; \\         
\>     $\eta = \eta \cap \bigcap_{v \in M} ( F(u,p_{max}) -
R_{u\rightarrow v}$); \\   
$N(u) = A - \NonNbrNodes$; \\
$p(u) = \min\{p:F(u,p) \supseteq \eta\}$
\end{tabbing}
\caption{Algorithm SMECN running  at node $u$.
\label{fig-SMECN}
}
\end{figure}

Define the graph $G=(V,E)$ by taking $(u,v) \in E$ iff $v \in N(u)$, as
constructed by the algorithm in Figure~\ref{fig-SMECN}.  It is immediate
from the earlier discussion that $E \supseteq E_2$. Thus
\begin{theorem} 
\label{the-SMECNk2}
$G$ has the minimum-energy property.
\end{theorem}
\commentout{
\prf
%
It suffices to show that $E_2$ consists of all the edges in $E'$ that
are not 2-redundant. 
$\forall (u,v)$ $\in$ $E'$, if $(u,v)$ $\not \in$ $E_2$, i.e. $v \not
\in N(u)$, then $\Loc(v)$ $\not \in$ ${\eta}^*$ (otherwise, by definition of
${\eta}^*$, $v$ is not in the relay region of any node in $A^*$. since
the search region $F^*$ $\supseteq$ ${\eta}^*$, $v$ $\in$ $A^*$. Then
$v$ $\in$ $N(u)$ according to our invariant ${\imath}^*$).    
Since $(u,v)$ $\in$ $E'$, $d(u,v) \leq d_{max}$ and $v$ $\in$
$Tr(u,d_{max})$. Thus, $\Loc(v)$ $\in$ $\bar{{\eta}^*}$ $\bigcap$
$Tr(u,d_{max})$ $=$ $\bigcup_{w\in A^*(u)}
R_{u\rightarrow w}$ $\bigcap$ $Tr(u,d_{max})$. Since $v$ is in the
union of the relay region of $A^*(u)$, there must exist a node $w \in
A^*(u)$ such that $v \in R_{u->w}$. By definition of relay region,
$p(u,v)$ $<$ $C((u,w,v))$. Therefore, we have show that $\forall
(u,v)$ $\in$ $E'$, if $(u,v)$ $\not \in$ $E_2$, then $(u,v)$ is a
2-redundant edge, i.e. all the edges that are not 2-redundant are in
$E_2$.  
\eprf}



\begin{figure}[htb]
\begin{tabbing}
\underline{Algorithm MECN}   \\
$p = p_0$;\\
$A = \emptyset$; \\
$\NonNbrNodes = \emptyset;$ \\
$\eta = F(u,p_{max})$; \\
while \= $F(u,p) \not \supseteq \eta$ do\\
\>     $p = \Increase(p)$;\\
\>     Broadcast NDM with power $p$ and gather responses;\\
\>     $M = \{v | \Loc(v) \in F(u,p),  v \not \in A, v \not= u\}$; \\
\>     $A = A \bigcup M$; \\
\>     $\NonNbrNodes = \NonNbrNodes \bigcup M$; \\
\>     for \=each $v \in M$ do $\flipAllStatesDownChain(v)$; \\
\>     $\eta = \bigcap_{v \in (A-\NonNbrNodes)} (F(u,p_{max}) -
R_{u\rightarrow v})$; \\   
$N(u) = A - \NonNbrNodes$;  \\
$p(u) = \min\{p:F(u,p) \supseteq \eta\}$\\
\\
Procedure $\flipAllStatesDownChain(v)$ \\
\>     if  $v \not \in \NonNbrNodes$ then  \\
\>\>          $\NonNbrNodes = \NonNbrNodes \bigcup \{v\}$; \\
\>\>          for each \= $w \in A$ such that $\Loc(w) \in R_{u\rightarrow v}$ do  \\
\>\>\>                $\flipAllStatesDownChain(w)$; \\
\>     else if $\Loc(v) \notin \union_{w \in A - \NonNbrNodes}
R_{u\rightarrow w}$ then \\
\>\>          $\NonNbrNodes = \NonNbrNodes - \{v\}$; \\
\>\>          for each $w \in A$ such that $\Loc(w) \in R_{u\rightarrow v} $  do \\
\>\>\>                  $\flipAllStatesDownChain(w)$; \\

\end{tabbing}
\caption{Algorithm MECN running at node $u$.
\label{fig-MECN}
}
\end{figure}
We next show that SMECN dominates MECN. MECN is described in
Figure~\ref{fig-MECN}. 
For easier comparison, 
we have made some inessential changes to MECN to make the notation and
presentation more like that of SMECN.
The main difference between SMECN and MECN is 
the computation of the region
$\eta$. 
As we observed, in SMECN, $\eta =  \cap_{v \in A}  (F(u,p_{max})
-R_{u\rightarrow v})$ at the end of every iteration of the loop.  On the
other hand, in MECN,
$\eta =  \cap_{v \in A - \NonNbrNodes}  (F(u,p_{max}) -R_{u\rightarrow
v})$.  Moreover, in SMECN, a node is never removed from
$\NonNbrNodes$ once it is in the set, while in MECN,  
it is possible for a node to be removed from $\NonNbrNodes$
by the procedure
$\flipAllStatesDownChain$. 
Roughly speaking,
if a node $v \in R_{u\rightarrow w}$, then, in the
next iteration, if $w \in R_{u\rightarrow t}$ for a newly
discovered node $t$, but $v \notin R_{ u\rightarrow t}$, 
node $v$ will be removed from $\NonNbrNodes$
by $\flipAllStatesDownChain(v)$.
In \cite{Rodoplu99}, it is shown that MECN is correct (i.e., it
computes a graph with the minimum-energy property) and terminates (and, in
particular, the procedure $\flip$ terminates).  Here we show that, at
least for circular search regions,  SMECN does better than MECN.

\begin{theorem}\label{MECNvsSMECN} 
If the 
search regions considered by the algorithm SMECN are circular, then
the communication graph 
constructed by SMECN
is a subgraph of the communication graph constructed by MECN.
\end{theorem}
\prf 
For each variable $x$ that appears in SMECN, let $x_S^k$ denote the
value of $x$ after the $k$th iteration of the loop; similarly, for each
variable in MECN, let $x_M^k$ denote the
value of $x$ after the $k$th iteration of the loop.
It is almost immediate
that SMECN maintains the following invariant:
$v$ $\in$ $\NonNbrNodes^k_S$ iff  $v$ $\in$ $A^k_S$ and $\Loc(v)$ $\in$
$\cup_{w \in A^k_S} R_{u\rightarrow w}$.  Similarly, it is not hard to
show that  
MECN maintains the following invariant:
$v$ $\in$ $\NonNbrNodes^k_M$ iff $v$ $\in$ $A^k_S$ and $\Loc(v)$ $\in$
$\cup_{w \in A^k_M - \NonNbrNodes^k_M} R_{u\rightarrow w}$.  
(Indeed, the whole point of the $\flip$ procedure is to maintain this
invariant.)  
Since it is easy to check that $A^k_S$ $=$ $A^k_M$, it is immediate 
that $\NonNbrNodes^k_S$ $\supseteq$ $\NonNbrNodes^k_M$.
Suppose that SMECN terminates after $k_S$ iterations of the loop and
MECN terminates after $k_M$ MECN iterations of the loop.
Hence $\eta^k_S$ $\subseteq$ $\eta^k_M$ for all $k$ $\le$ $\min(k_S,k_M)$.
Since both algorithms use the condition $F(u,p) \supseteq \eta$ to
determine termination, it follows that 
SMECN terminates no later than MECN; that is, $k_S \le k_M$.

Since the search region used by SMECN is assumed to be circular,
by Proposition~\ref{N2char}, $A_S^{k_S}-\NonNbrNodes_S^{k_S}
= N_2(u)$. 
Moreover, even if we continue to iterate the loop of SMECN (ignoring the
termination condition), then $F(u,p)$ keeps increasing while $\eta$
keeps decreasing.  Thus, by Proposition~\ref{N2char} again, 
we continue to have $A_S^k-\NonNbrNodes_S^k = N_2(u)$ even if $k \ge k_S$.
That means that if we were to continue with the loop after SMECN
terminates, none of the new nodes discovered would be neighbors of $u$.
Since the previous argument still applies to show that
$\NonNbrNodes_S^{k_M} \supseteq \NonNbrNodes_M^{k_M}$, it follows that 
$N_2(u) = A_S^{k_M} - \NonNbrNodes_S^{k_M} \subseteq A_M^{k_M} -
\NonNbrNodes_M^{k_M}$.  That is, the communication graph constructed by SMECN
has a subset of the edges of the communication graph constructed by
MECN.
\eprf


In the proof of Theorem~\ref{MECNvsSMECN}, we implicitly assumed that
both SMECN and MECN use the same value of initial value $p_0$ of $p$ 
and the same function $\Increase$.  In fact, this assumption is not
necessary, since the neighbors of $u$ in the graph computed by SMECN are
given by $N_2(u)$ independent of the choice of $p_0$ and
$\Increase$, as long as $F(u,p_0) \not\supseteq F(u,p_{max})$ and
$\Increase^k(p_0) \ge p_{max}$ for $k$ sufficiently large.  Similarly,
the proof of Theorem~\ref{MECNvsSMECN} shows that the set of neighbors of $u$
computed by MECN is a superset of $N_2(u)$, as long as $\Increase$
and $p_0$ satisfy these assumptions.

Theorem~\ref{MECNvsSMECN} shows that the neighbor set computed by MECN
is a superset of $N_2(u)$.  As the following example shows, it may be a
strict superset (so that the communication graph computed by SMECN is a
strict subgraph of that computed by MECN).

\begin{example}
Consider a network with 4 nodes $t, u, v, w$, where
$\Loc(v) \in R_{u   \rightarrow w}$, $\Loc(w) \in R_{u \rightarrow t}$,
and $\Loc(v) \notin R_{u \rightarrow t}$.  It is not hard to choose
power functions and locations for the nodes which have this property.
It follows that $N_2(u) = \{t\}$.  (It is easy to check that 
$\Loc(t) \notin R_{u   \rightarrow v} \cup R_{u \rightarrow w}$.)
On the other hand, suppose that $\Increase$ is such that $t$, $v$, and
$w$ are added to $A$ in the same step.  Then all of them are added to
$\NonNbrNodes$ in MECN.  Which ones are taken out by $\flip$ then
depends on the order in which they are considered in the loop
\footnote{Note that the final neighbor set of MECN is 
claimed to be
  independent of the ordering in \cite{Rodoplu99}. However, 
the example here shows that this is not the case.}
For example, if they are considered in the order $v$, $w$, $t$, then the
only neighbor of $u$ is again $t$.  However, if they are considered in
any other order, then both $v$ and $t$ become  neighbors of $u$.  For
example, suppose that they are considered in the order $t$, $w$, $v$.
Then $\flip$ makes $t$ a neighbor, does not make $w$ a neighbor (since
$\Loc(w) \in R_{u \rightarrow t}$), but does make $v$ a neighbor (since
$\Loc(v) \notin R_{u \rightarrow t}$).  Although 
$\Loc(v) \notin R_{u \rightarrow w}$, this is not taken into account
since $w  \in \NonNbrNodes$ at the point when $v$ is considered.
\end{example}

\commentout{
$v, w$ $\not \in$ $N(u)$. Since $t \not \in R_{u
 \rightarrow v}$ and $t \not \in R_{u
 \rightarrow w}$, $t \in N(u)$. Therefore $N(u)=\{t\}$. For MECN,
 $w$ $\in$ $R_{u \rightarrow t}$ and  $v$ $\not \in$ $R_{u
 \rightarrow t}$, so $v$ $\in$ $N(u)$. In addition,$t$ $\in$ $N(u)$.  
 Therefore, $\overline{N(u)} = \{t,v\}$. Similarly, $N(v)$ $=$
 $\overline{N(v)}$ $=$ $\{w\}$, $N(w)$ 
 $=$ $\overline{N(w)}$ $=$ $\{v,t\}$ and $N(t)$ $=$ $\overline{N(t)}$
 $=$ $\{u,w\}$. Therefore, MECN maintains one more edge $(u,v)$. Note
 that the edge is asymmetric, i.e. $v \in \overline{N(u)}$, but $u \not \in
 \overline{N(v)}$. In contrast, all edges in SMECN are symmetric if
 the final search region is circular.   
\eprf
\input{epsf}
\begin{figure*}[ht]
\begin{center}
\begin{tabular}{cc}
\epsfysize=6.0cm \epsffile{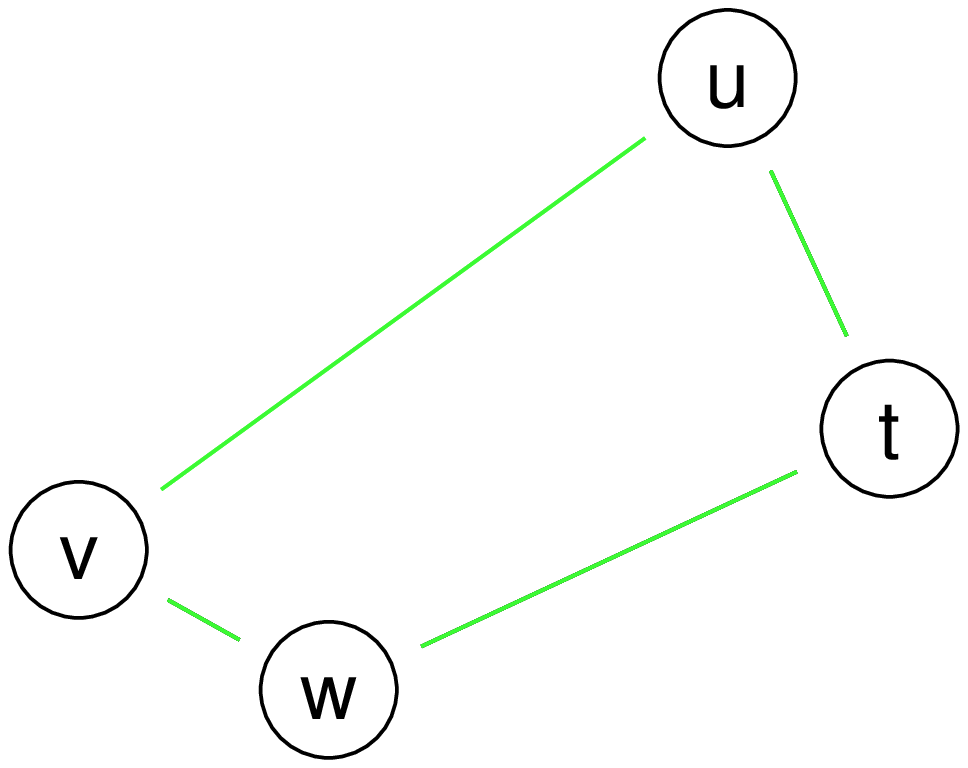}    &
\epsfysize=6.0cm \epsffile{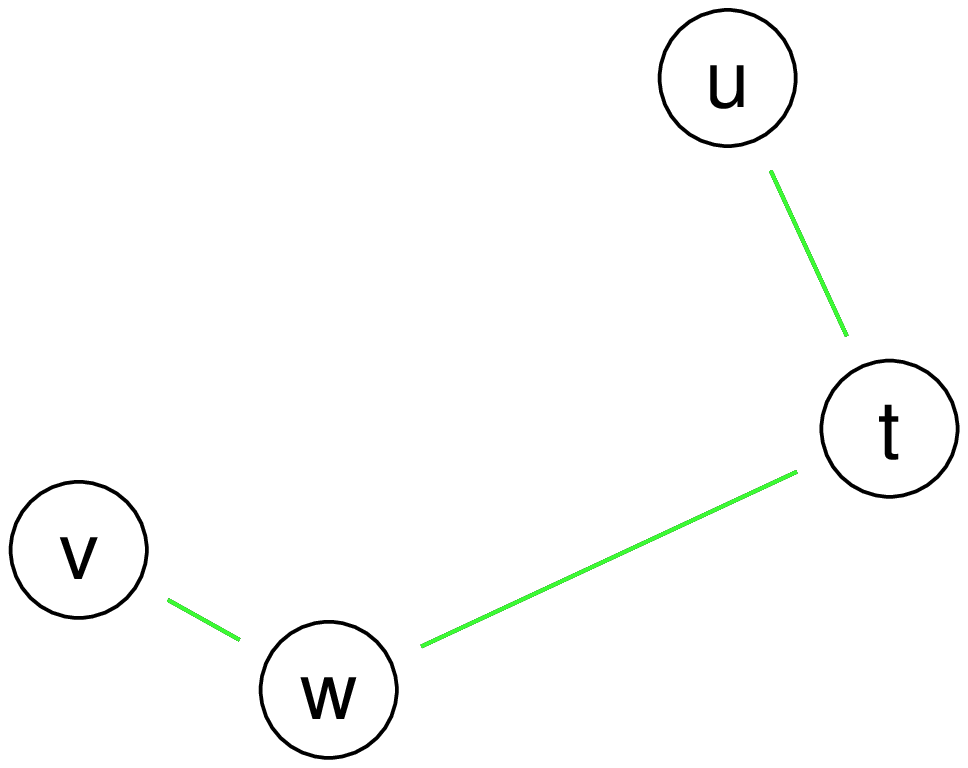}  \\
{\footnotesize (a) MECN}   &
{\footnotesize (b) SMECN}   
\end{tabular}
\end{center}
\caption{
A network where SMECN dominates MECN.
\label{fig-nd4ex}
}
\end{figure*}
%

}

\commentout{
\section{Discussions 
}
\label{sec-discussions}

In Section \ref{sec-SMECN}, we have assumed that the broadcast of NDM
is reliable. Here we discuss what to do if we have an unreliable MAC
layer. In addition, we discuss how to maintain the minimum energy
property of the communication graph $G$ if the network topology
changes dynamically over time due to mobility or failure. 

\subsection{
Dealing with an unreliable MAC broadcast
}
\commentout{
The Neighbor discovery broadcast message (NDM) can be sent by the Neighbor
Discovery Protocol (NDP). A NDP is needed by many routing protocols
such as ZRP \cite{zrp-03} and GPSR \cite{GPSR00}. Since our protocol is
envisioned to provide support for the routing protocol, we can use the NDP
available to the routing protocol. A NDP is usually a simple beaconing
algorithm to provide all nodes with their neighbors' information
(addresses, positions, etc). In our case, each node periodically
transmits a beacon to the broadcast MAC address with its own addresses
and position. In order for the receiver to send a prompt beacon back,
we need a special flag in the beacon. Let's call it the URGENT
flag. It is well-known that periodic messages can lead to adverse
synchronization effects \cite{Floyd94}. To avoid this problem, each
regular 
beacon's transmission needs to jitter by 50\% of the mean beacon
interval and each URGENT beacon's transmission needs to jitter a small
random time. This small random time is usually uniformly distributed
in a small interval such as [0,10] milliseconds. We show that with a
NDP, algorithm SMECN quickly converges to its ideal neighbor set (the
neighbor set obtained with reliable MAC broadcast).  
} 

A Neighbor Discovery Protocol (NDP) is usually a simple periodic
beaconing algorithm to provide all nodes with their neighbors'
information (addresses, positions, etc). A NDP is needed by many
routing protocols such as ZRP \cite{zrp-03}, GPSR \cite{GPSR00}. Since
our protocol is meant to provide support for the routing protocol, we
can use the NDP to send the Neighbor discovery broadcast message
(NDM).
 In our case, each node periodically
transmit a beacon to the broadcast MAC address with its own addresses
and position. In order for the receiver to send a prompt beacon back
(i.e. the NDM response message),
we need a special flag in the beacon. Let's call it the URGENT
flag. It is well-known that periodic messages can lead to adverse
synchronization effects \cite{Floyd94}. To avoid this, each regular
beacon's transmission needs to jitter by 50\% of the mean beacon
interval. Each URGENT beacon's transmission needs to jitter a small
random time. This small random time is usually uniformly distributed
in a small interval such as [0,10] milliseconds.  
We show that with a NDP, algorithm SMECN converges to its ideal
neighbor set (the neighbor set obtained using a reliable MAC
broadcast).

\begin{theorem} 
Define ${p_a}^*$ to be the power $p$ in Figure~\ref{fig-SMECN} when
SMECN using an unreliable MAC broadcast terminates. If the NDP uses
the ${p_a}^*$ to send regular
beacons, then $N_a(u)$ converges to a set $N_b(u)$ such that
$N_b(u) \supseteq N_2(u)$. If the final search region ${F_a}^*$ is
circular, then $N_b(u)=N_2(u)$.
\end{theorem}
\prf 
Suppose SMECN using unreliable broadcast (denote it $SMCN_a$) is
running at the same time as SMECN using reliable broadcast. We
distinguish the variables in the former from those in the latter using
the subscript ``a''. Since 
they use the same $Increase(p)$ and initial $p$, and NDM messages and
responses can be lost for $SMCN_a$, $A_a$ $\subseteq$
$A$ at every iteration before either $SMCN_a$ or SMECN
terminates. Since $A_a$ $\subseteq$ $A$, ${\eta}_a$ $\supseteq$
$\eta$. Therefore, $SMCN$ terminates no later than
${SMECN}_a$. Therefore, ${p_a}^*$ $\ge$ $p^*$. Thus, if a node $v \in
A$, but $v \not \in A_a$, then node $v$ must be inside the final
search region ${F_a}^*$. Since beaconing with ${p_a}^*$ reaches any node
in ${F_a}^*$, $v$ will be discovered eventually (we assume messages
are received infinitely many times if they are sent infinitely many
times). After discovering a new node, node $u$ will recompute its
neighbor set (details will be discussed in the next subsection). By
Proposition~\ref{N2char}, $N_b(u) \supseteq N_2(u)$ after all the
nodes in $A$ are discovered. If the final search region ${F_a}^*$ is
circular, then $N_b(u)=N_2(u)$. 
\eprf
}

\commentout{
\section{Reconfiguration}\label{sec-reconfig}
In a multi-hop wireless network, nodes can be mobile. Even if nodes do
not move, nodes may die if they run out of energy. 
In addition, new nodes may be added to the network.
We assume that each node uses a Neighbor Discovery
Protocol (NDP), a periodic message that provides all its neighbors with
its current position (according to the GPS) in order to detect changes
in the topology of the network.  A node $u$ sends out the
message with just enough power to reach all the nodes that it currently
considers to be its neighbors (i.e., the nodes in $N_2(u)$).
Once a node detects a change, it may need to update its set of neighbors.
This is done by a {\em reconfiguration protocol}.
Rodoplu and Meng \cite{Rodoplu99}
do not provide an explicit reconfiguration protocol.
Rather, they deal with changes in network topology by running MECN
periodically at every node.  While this will work, it
is inefficient.  If a node does 
not detect any changes, then there is no obvious need to run MECN.
We now present a reconfiguration protocol where, in a precise sense,
we run SMECN  only when necessary  (in the sense that it is run only
when not running it may result in a network that 
does not satisfy the minimum-energy property).

There are three types of events that 
trigger the reconfiguration protocol: {\em leave events}, {\em join
events},  and
{\em move events}:
\begin{itemize} 
\item A $\leave_u(v)$ event happens when a node $v$ that was in $u$'s
neighborhood is detected to no longer be in the neighborhood (since its
beaconing message is  not received).  This may happen because $v$ 
is faulty or dies or because it has in
fact moved away.
\item A $\join_u(v)$ event
happens when a node $v$ is detected to be within $u$'s neighborhood by
the NDP. 
\item A $\move_u(v,L)$ event happens when $u$ detects that $v$ has
moved from the previous location to the current location $L$.  ($v$'s
location $L$ is 
relative
to $u$'s location, so the event could be
due to $u$'s own movement.)  
\end{itemize}


%

It is straightforward to see how to update the neighbor set if $u$
detects a single change.  
Suppose $p^*$ is $u$'s current power setting (that is, the final
power setting used in the last invocation of SMECN by $u$);  let $F^*
= F(u,p^*)$ be the last region searched by $u$.
let $A^*$ consist of all the nodes in $F^*$ (that is, the set of all
nodes discovered by the algorithm).
\begin{itemize}
\item If a single $\leave_u(v)$ or a $\move_u(v,L)$ is detected, let
  $A' = A^* - \{v\}$ if $\leave_u(v)$ is detected and let $A' = A^*$ if
  $\move(v,L)$ is detected.   Let $R_F' = \bigcap_{w \in A'}
  (F(u,p_{max}) - R_{u\rightarrow w})$, 
where the new location for $v$ is used in the computation if $v \in A'$.
(Note that  $R_F'$ is defined essentially in the same way as $R_F(u)$ in
Equation (\ref{eq-RF}).)  If $F^* \supseteq R_F'$, then take $u$'s updated neighbor
set to be $N_{R_F'}$; otherwise, run SMECN taking $p_0 = p^*$.  
\item If a single $\join_u(v)$ is detected, recompute the neighbor set
as follows.
Let $A' = A^* \cup \{v\}$.
Let $R_F' = \bigcap_{w \in A'} (F(u,p_{max}) - R_{u\rightarrow w})$.
Take $u$'s updated neighbor set to be $N_{N_{R_F'}}$.  Then let $p' = 
\min\{p: F(u,p) \supseteq  \bigcap_{w \in A'} (F(u,p_{max}) -
R_{u\rightarrow w})\}$. 
\end{itemize}

The following proposition is almost immediate from our earlier work.
\begin{proposition} Suppose that a graph $G$ has the minimum-energy
property.  If the nodes in $G$ observe a sequence of single changes and update
their edge sets as above, the resulting graph $G^*(V,E^*)$ still has the
minimum-energy property for the new topology.  Moreover, if $F(u,p)$ is
a circular region for all $p$, then $E^* = E_2$. 
\end{proposition}

In general, there may be more than one change event that is detected at a
given time by a node $u$. (For example, if $u$ moves, then there will in
general 
several leave and move events detected by $u$.)  If more than one change
event is detected by $u$, we
consider the events observed in some order.  If we can perform all the
updates without rerunning SMECN, we do so; otherwise, we rerun SMECN
starting from $p^*$.  By rerunning SMECN, we can deal with all the
changes simultaneously.

\commentout{
\begin{theorem} 
\label{the-leave}
For a single $\leave_u(v)$ event, let ${\hat{R}}_{F^*}(u)$ be computed
using $N_{F^*} - \{v\}$ in Equation~\ref{eq-RF}. If $v \in N(u)$ and
$F^* \not \supseteq {\hat{R}}_{F^*}(u)$, then $u$ need to rerun
algorithm SMECN with $p$ initialized to be $p^*$. Otherwise, $u$ only
need to recompute its neighbor set, i.e. the neighbor set need to be
updated to $N_{{\hat{R}}_{F^*}(u)}$.
\end{theorem}
\prf
   Only $u$'s neighbor set can be affected by the $\leave_u(v)$ event.
If $v \in N(u)$ and $F^* \supseteq {\hat{R}}_{F^*}(u)$, recompute
$u$'s neighbor set, i.e. $\hat{N}(u) = N_{\hat{R}_{F^*}(u)}$. By
Proposition~\ref{N2char}, $\hat{N}(u) \supseteq \hat{N}_2(u)$.  If $v
\in N(u)$ and $F^* \not \supseteq \hat{R}_{F^*}(u)$, then $u$ will
rerun Algorithm SMECN. By Proposition~\ref{N2char}, $\hat{N}(u)
\supseteq \hat{N}_2(u)$.  Therefore, after the $\leave_u(v)$ event, the
constructed graph $\hat{G}=(V,\hat{E})$ maintains the minimum-energy
property.  If the nodes we require to rerun the algorithm does 
not run it, then the new communication network may not satisfy the
minimum energy property or is disconnect connected. For example, in
Figure~\ref{fig-nd4ex}-(b), assume $v$ and $t$ can reach each other
using maximum power. If node $w$ dies, and $v$ and $t$ do not rerun
the algorithm, then $v$ and $t$ cannot communicate with each other.
\eprf

\begin{theorem} 
\label{the-join}
For a single $\join_u(v)$ event, only $u$ needs to recompute its
neighbor set (There is no need to send NDM).
\end{theorem}
\prf  
Only $u$'s neighbor set can be affected by the $\join_u(v)$
event. Since ${F_u}^* \supseteq R_{F^*}(u)$, ${F_u}^* \supseteq
\hat{R}_{F^*}(u)$ where $\hat{R}_{F^*(u)}$ is computed using $N_{F^*}
+ \{v\}$ in Equation~\ref{eq-RF}.  By Proposition~\ref{N2char},
$\hat{N}(u) \supseteq \hat{N}_2(u)$.  Therefore, after the $\join_u(v)$
event, the constructed graph $\hat{G}=(V,\hat{E})$ maintains the
minimum energy property. Note that, in order to conserve NDP beacon
power,  $u$ can choose to decrease $p^*$ to the minimum ${\hat{p}}^*$
such that $F(u,{\hat{p}}^*)$ $\supseteq$ $\hat{R}_{F^*}(u)$.   
\eprf

\begin{theorem} 
\label{the-move}
For a single $\move_u(v,L)$ event, if $F^* \not \supseteq
\hat{R}_{F^*}(u)$ where $\hat{R}_{F^*(u)}$ is computed using $v$'s new
location, then $u$ need to rerun algorithm SMECN with $p$
initialized to be $p^*$. Otherwise, $u$ only needs to recompute its
neighbor set, i.e. the neighbor set needs to be updated to
$N_{\hat{R}_{F^*}(u)}$.
\end{theorem}
\prf   
The proof is similar to that for Theorem~\ref{the-leave}.  
For the case where $u$ only need to recompute its neighbor set, in order
to conserve NDP beacon power,  $u$ can choose to decrease $p^*$
to the minimum ${\hat{p}}^*$ such that $F(u,{\hat{p}}^*)$ $\supseteq$
$\hat{R}_{F^*}(u)$.   
\eprf
}

\commentout{
\begin{theorem} 
\label{the-combine}
For any node $u$, if there are several other events happening before
an event $e_u$ of node $u$ finishes ($e_u$ can  only be an event that
involves sending NDM messages. We assume that recomputation of a
node's neighborhood cannot be interrupted before its completion.),
then events can be combined as follows: (1) If a $\leave_u(v)$ event
happens, $u$ needs to remove $v$ from $A$. (2)  If a $\join_u(v)$ event
happens, then $u$ simply adds $v$ to $A$.  (3) If a $\move_u(v,L)$
event happens, $u$ needs to update $v$'s location. For all the three
types of events, recompute the current $\NonNbrNodes$ and $\eta$ in
Figure~\ref{fig-SMECN}. If the current search region $F$ $\not
\supseteq$ $R_{F}(u)$, then $u$ should continue its ``while''
loop. Otherwise, $u$ can terminate the ``while'' loop. The above
process ensures the minimum energy property of the current network
topology.
\end{theorem}
\prf   
Let $G'=(V,E')$ be the previous graph where $E'= \{ (u,v) |$,
$d(u,v)$ $\leq$ $d_{max}\}$. Let $\hat{G}=(V,\hat{E})$ be the new
graph where $\hat{E} = \{ (u,v) |$, $d(u,v)$ $\leq$ $d_{max}\}$
after the topology changes (The series of events are triggered by
these 
topology changes). To show that the constructed graph
$\hat{G}=(V,\hat{E})$ satisfies the minimum energy property in
$\hat{G'}$, all we need to show is that each node $u$ constructs
$\hat{N}_2(u)$, where $v \in \hat{N}_2(u)$ if $(v,u) \in \hat{G}$.  
Now assume there is no topology changes after the series of events.
Since (1)-(3) provides all the up-to-date information on nodes and
their positions in current search region, by Proposition~\ref{N2char},
$\hat{N}(u) \supseteq \hat{N}_2(u)$ after node $u$ terminates SMECN
algorithm. 
\eprf

\begin{corollary} 
If the final search region is circular, the reconfiguration algorithm
converges to the up-to-date graph $E_2$. (It will construct
the up-to-date $E_2$ if there is no topology changes, i.e. no other
events will happen in the future).
\end{corollary}
}

Up to now we have assumed that no topology changes are detected while
SMECN itself is being run.  If changes are in fact detected while SMECN
is run, then it is straightforward to incorporate the update into SMECN.
For example, if $u$ detects a $\join_u(v)$ event, then $v$ is added to
the set $A$ in the algorithm, while if $u$ detects a $\leave_u(v)$
event, $u$ is dropped from $A$ and $\eta$ is recomputed.  We leave the
details to the reader.


As we mentioned earlier, there is no reconfiguration protocol given in
\cite{Rodoplu99}.  However, it is easy to modify the reconfiguration
algorithm protocol given above for SMECN so that it works for MECN.
If a $\leave_u(v)$ or $\move_u(v,L)$ is detected, then the same approach
works (except that MECN rather than SMECN is called with $p_0 = p^*$).
Similarly, if a $\join_u(v)$ is detected, we update the neighbor set
using the approach of MECN rather than SMECN.

Note that we have assumed a perfect MAC layer in our reconfiguration
discussion. Our reconfiguration works fine 
even with a MAC layer that drops
packets. The reason is as follows. If the NDM responses of some nodes
get dropped, then the final power setting ${p_a}^*$ using an imperfect
MAC layer will be bigger than the corresponding $p^*$ using a perfect MAC
layer. Since NDP beaconing with $p^*$ reaches all nodes in $N_2(u)$,
beaconing with a bigger power ${p_a}^*$ will still reach all nodes
in $N_2(u)$. Eventually all the nodes in $N_2(u)$ whose responses are
lost will be detected by $u$ through NDP beacons. Thus, the neighbor
set computed using an imperfect MAC layer converges to a superset of
$N_2(u)$. If the final search region is circular, then the neighbor
set converges to the set $N_2(u)$.
} 

\section{Simulation Results and Evaluation} 
\label{sec-sim}
How can using the subnetwork computed by (S)MECN help performance?
Clearly, sending messages on minimum-energy paths is more efficient than
sending messages on arbitrary paths, but the algorithms are all local;
that is, they do not actually find the minimum-energy path, they just
construct a subnetwork in which it is guaranteed to exist.

There are actually two ways that the subnetwork constructed by (S)MECN
helps.  First, when sending periodic beaconing messages, it suffices for
$u$ to use power $p(u)$, the final power computed by (S)MECN.  Second,
the routing algorithm is restricted to using the edges $\cup_{u \in
V} N(u)$.  While this does not guarantee that a minimum-energy path is
used, it makes it more likely that the path used is one that requires
less energy consumption.

To measure the 
effect of
focusing on energy efficiency, we compared the use of MECN and SMECN in
a simulated application setting.  

Both SMECN and MECN were implemented in ns-2 \cite{ns_2},
using the wireless extension developed at Carnegie Mellon \cite{cmu-nse}.
The simulation was done for a network of 200 nodes,
each with a transmission range of 500 meters.
The nodes were placed uniformly at random in a rectangular
region of 1500 by 1500 meters. (There has been a great deal of work on
realistic placement, e.g. \cite{Zegura96,Zegura97}. However, 
this
work has the Internet in mind. 
Since the nodes in a multihop network are often best viewed as being
deployed in a somewhat random fashion and move randomly, 
we believe that the uniform
random placement assumption is reasonable in many large multihop wireless
networks.) 

We assume a $1/d^4$ transmit power roll-off for radio propagation.
The carrier frequency is 914 MHz;
transmission raw bandwidth is 2 MHz.
We further assume that each node has an omni-directional antenna with 0 dB
gain, which is placed 1.5 meter above the node. 
The receive threshold is 94 dBW,
the carrier sense threshold is 108 dBW, and the capture threshold is
10 dB. These parameters simulate the 914 MHz Lucent WaveLAN DSSS
radio interface.
Given these parameters, the $t$ parameter in
Section~\ref{sec-model} is 101 dBW. 
We ignore reception power consumption, i.e. $c=0$.



Each node in our simulation has an initial energy of 1
Joule.  We would like to see how our algorithm affects network
performance. To do this, we need to simulate the network's application
traffic.
We used the following application scenario. All nodes periodically
send UDP traffic to
a sink node situated at the boundary of the network.  
The sink node is
viewed as the master data collection site.  
The application traffic is assumed to be  CBR
(constant bit rate); application packets are all 512 bytes.
The sending rate is 0.5 packets per second.
%
This application scenario has also been used in
\cite{Heinzelman00}. 
Although this application scenario does not seem appropriate for
telephone networks and the Internet (cf.~\cite{PaxFlo95,PaxFlo97}), it
does seem reasonable for ad hoc networks, for example, in
environment-monitoring sensor applications.  In this setting, sensors
periodically  transmit data to a data collection site, where the data is
analyzed. 

To find routes along which to send messages, we use AODV~\cite{AODV99}.
However, as mentioned above, we restrict AODV to finding routes that use 
only edges in $\cup_v N(u)$.
There are other routing protocols, such as LAR \cite{LAR98},
GSPR \cite{GPSR00}, and DREAM \cite{DREAM98}, that take advantage of GPS
hardware.  We used AODV because 
it is readily available in our simulator 
and it is well studied.  We do not believe that using a
different routing protocol would significantly affect the results we
present here.  

We assumed that each node in our simulation had an
initial energy of 1 Joule and then ran the simulation for 1200
simulation seconds, using both SMECN and MECN.
We did not actually simulate the execution of SMECN and MECN.  Rather,
we assumed the neighbor set $N(u)$ and power
$p(u)$ computed by (S)MECN each time it is run were given by an
oracle.
(Of course, it is easy to compute 
the neighbor set and power
in the simulation, since
we have a global picture of the network.)  Thus, in our simulation, we
did not take
into account one of the benefits of SMECN over MECN, that it stops
earlier in the neighbor-search process.
Since a node's available energy is
decreased after each packet reception or transmission, nodes in the
simulation die over time. 
After a node dies, the network must be reconfigured. 
In \cite{Rodoplu99}, this is done by running MECN periodically.  In the
full paper, we present a protocol that does this more efficiently.
In our simulations, we have used this protocol (and implemented an
analogous protocol for MECN).

For simplicity, we simulated only a static  network
(that is, we assumed that nodes did not move), although some of the
effects of mobility---that is, the triggering of the reconfiguration
protocol---can already be observed with node deaths.  

In this setting, we were interested in network lifetime, as measured by
two metrics: (1) the number of nodes still alive over time and (2) the
number of nodes 
still connected to the sink.   

\begin{figure}[htb]
\begin{center}
\begin{tabular}{cc}
\epsfysize=4.2cm \epsffile{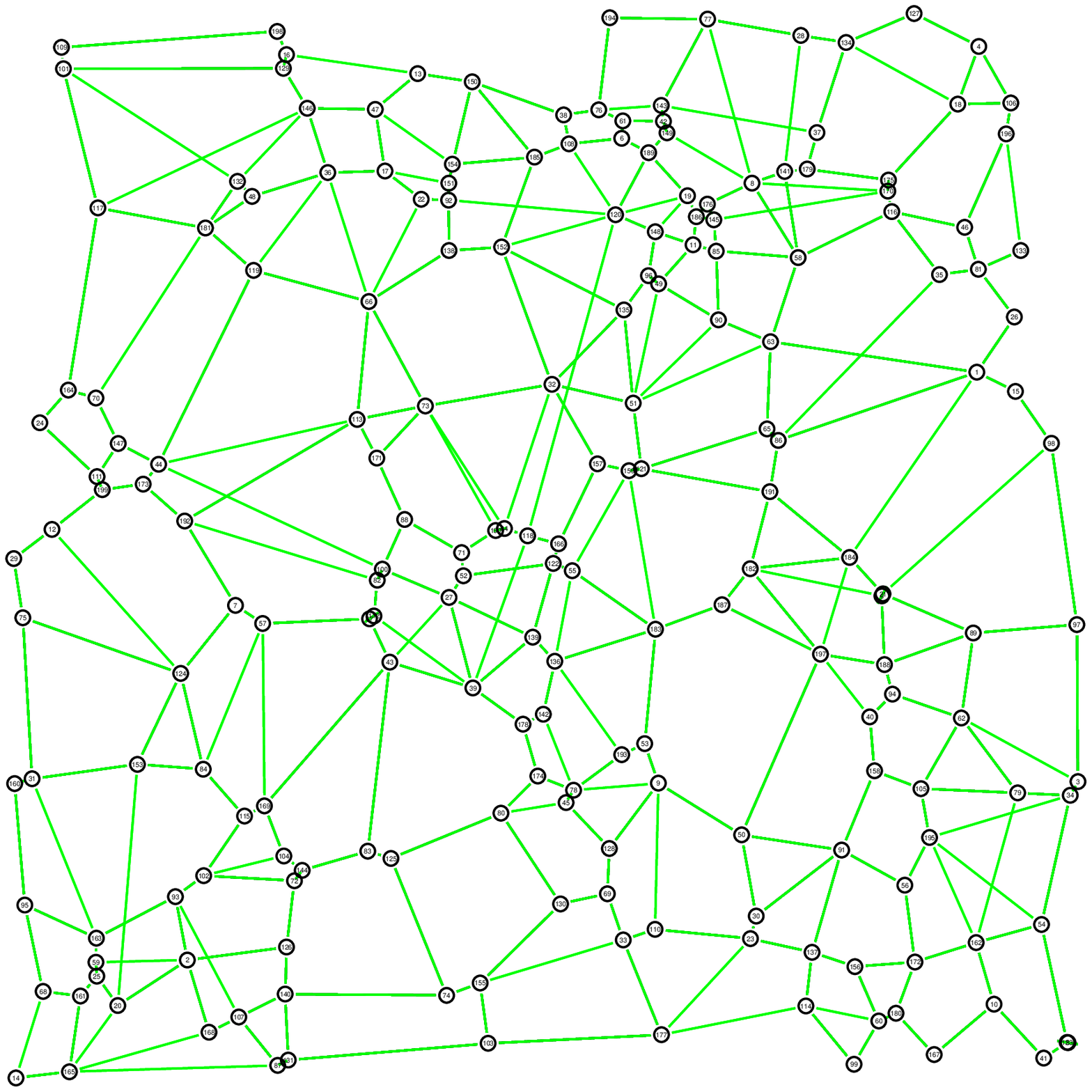}    &
\epsfysize=4.2cm \epsffile{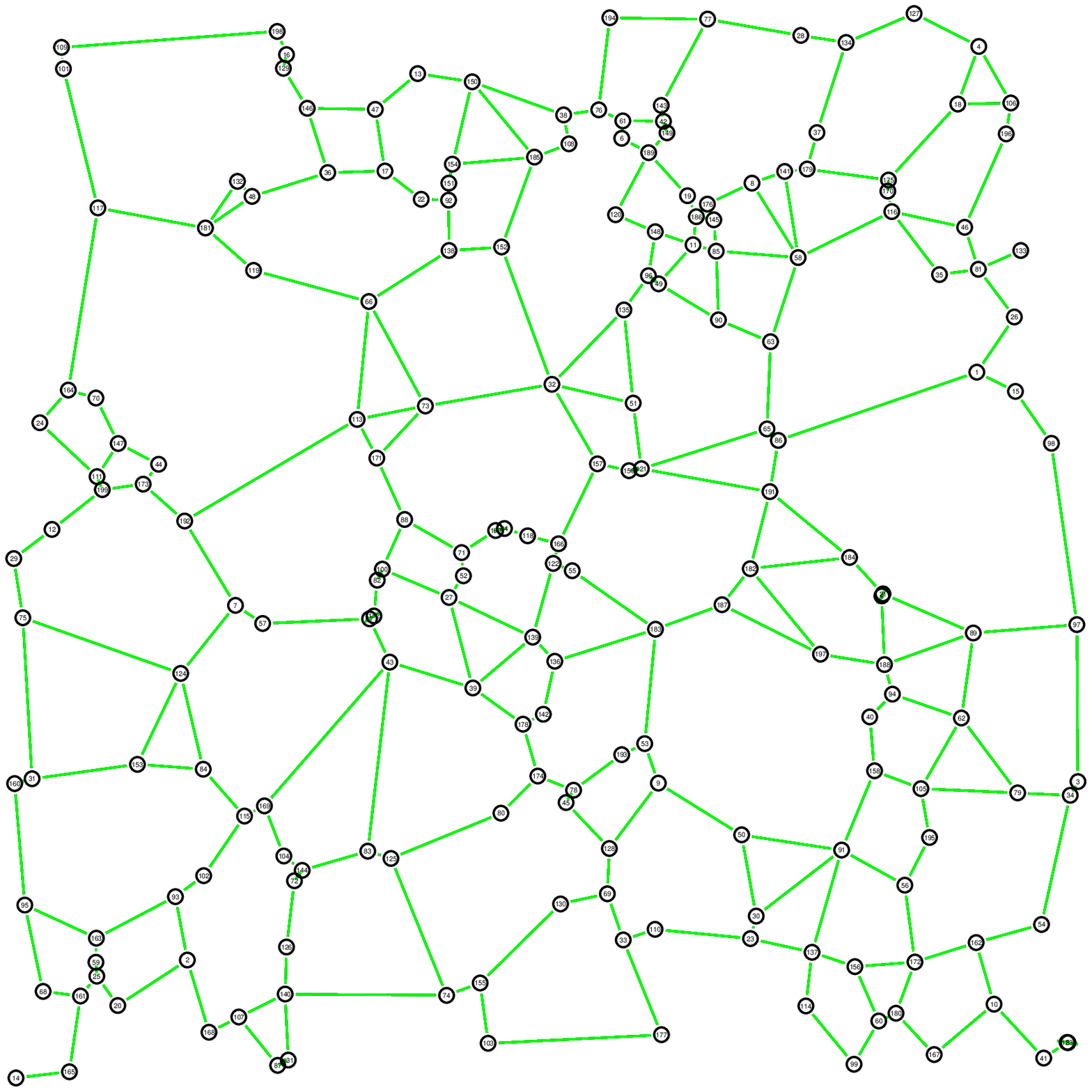}  \\
{\footnotesize (a) MECN}   &
{\footnotesize (b) SMECN}   
\end{tabular}
\end{center}
\caption{
Initial network computed by MECN and SMECN.
\label{fig-topographs}
}
\end{figure}

Before describing the performance, we consider
some features of the subnetworks computed by MECN and SMECN. 
Since the search regions will be circular with an omnidirectional
antenna, Theorem~\ref{MECNvsSMECN} assures us that the network used
by SMECN will be a subnetwork of that used by MECN, but it does not say
how much smaller it will be.  
The initial network 
in a typical execution of the MECN and SMECN is shown in Figure
\ref{fig-topographs}. The average number of neighbors of MECN and
SMECN are $3.64$ and $2.80$ respectively. 
Thus, each node running MECN has roughly 30\% more links than the same
node running SMECN.  This makes it likely that the final power setting
computed will be higher for MECN than for SMECN.  In fact, our
experiments show that it is roughly 49\% higher, so more
power will be used by nodes running  MECN when sending messages.
Moreover, AODV is unlikely
to find routes that are as energy efficient with MECN.
\input{epsf}

\begin{figure}[htb]
\begin{center}
\epsfysize=4.2cm \epsffile{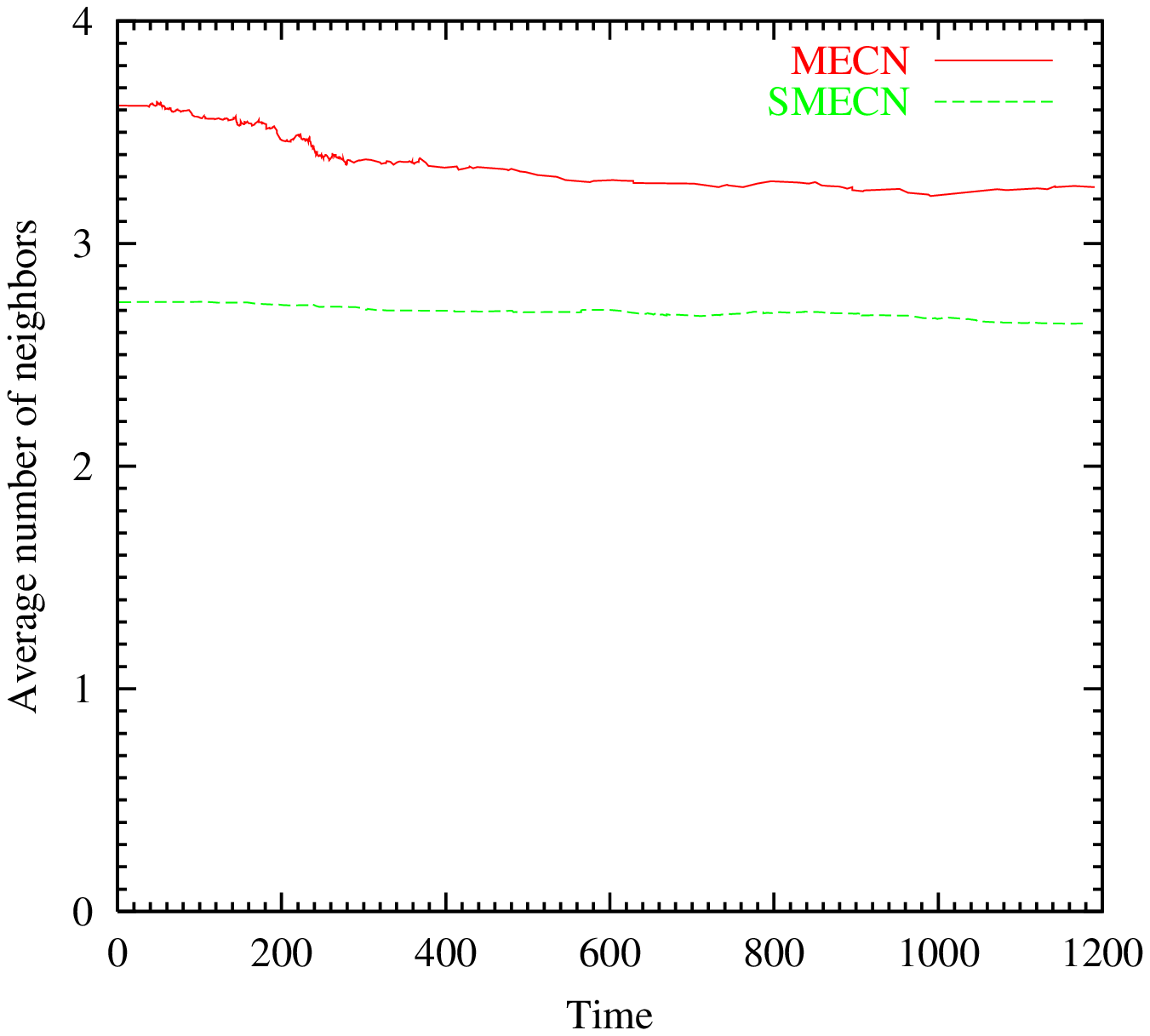} 
\end{center}
\caption{
Average number of neighbors over time. 
\label{fig-avgnbrs}
}
\end{figure}

As nodes die (due to running out of power), the network topology changes
due to reconfiguration.  Nevertheless, as shown in
Figure~\ref{fig-avgnbrs}, the average number of neighbors
stays roughly the same over time, thanks to
the reconfiguration protocol.

\input{epsf}

\begin{figure}[htb]
\setlength\tabcolsep{0.1pt}
\begin{center}
\begin{tabular}{cc}
\epsfysize=4.0cm \epsffile{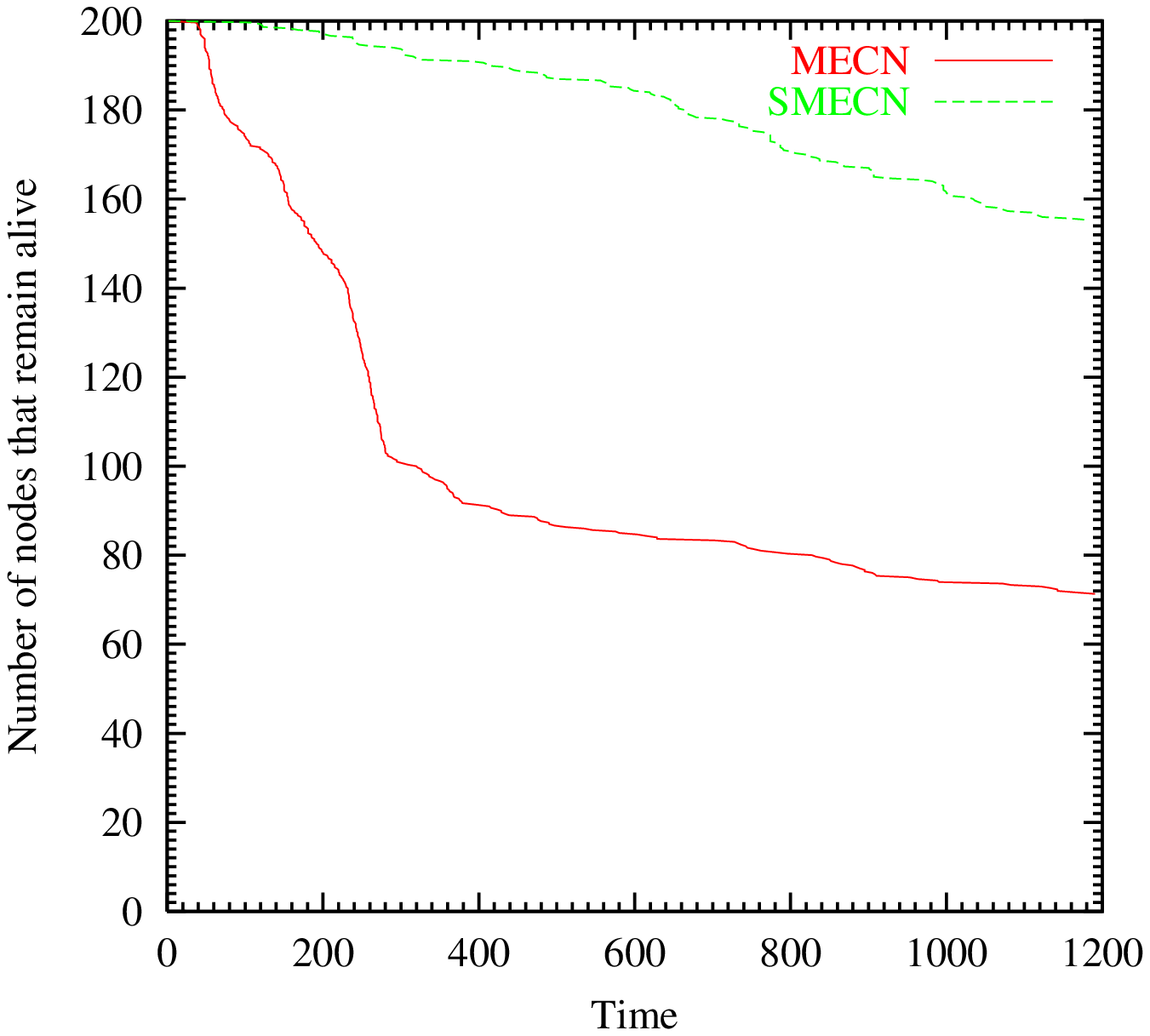}    &
\epsfysize=4.0cm \epsffile{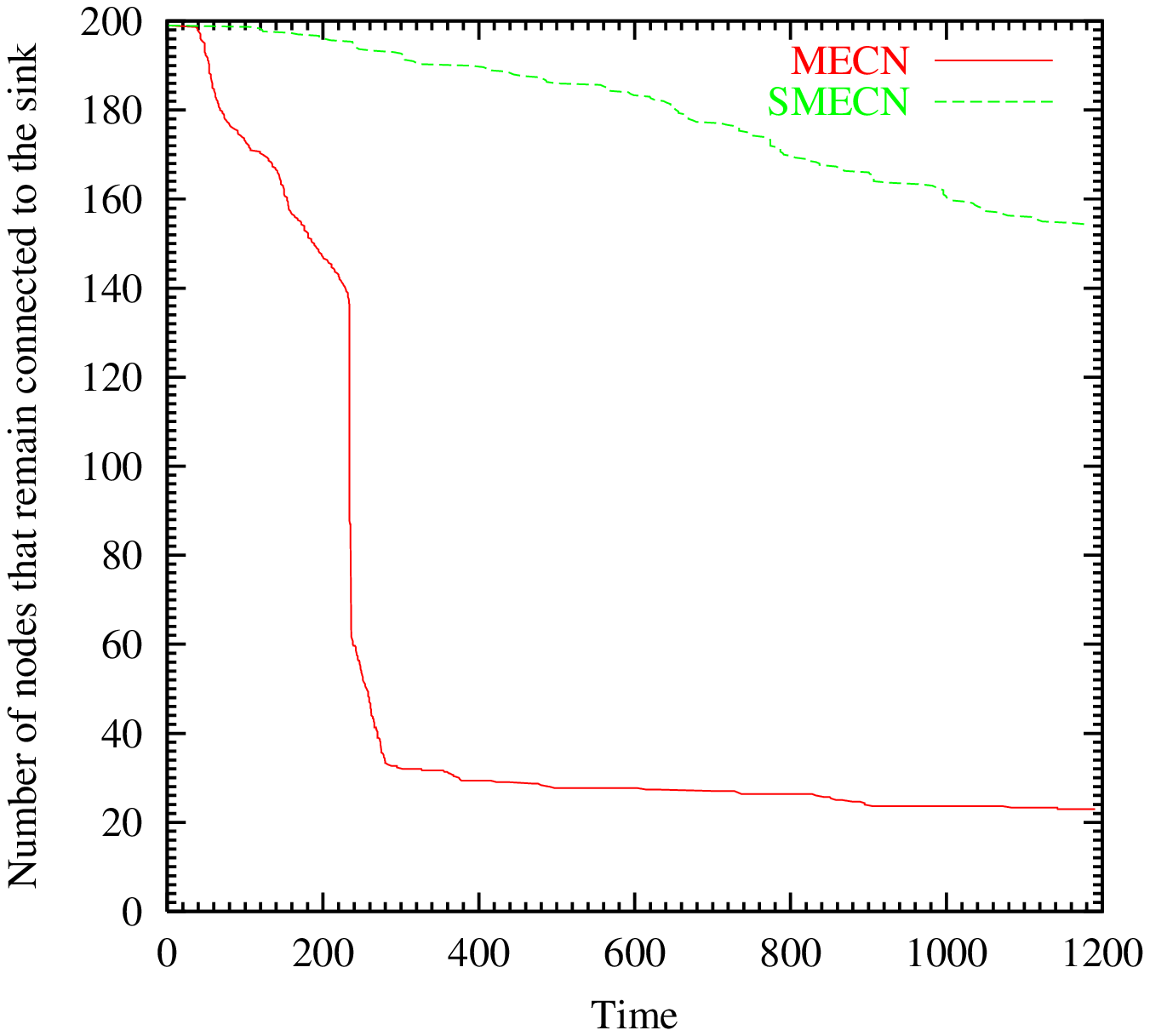}  \\
{\footnotesize (a) Number of nodes that remain alive}   &
{\footnotesize (b) Number of nodes that remain} \\
 &
{\footnotesize connected to the sink}   

\end{tabular}
\end{center}
\caption{
Network life time with respect to two different metrics. 
\label{fig-lifetime}
}
\end{figure}

Turning to the network-lifetime metrics discussed above, 
as shown in Figure \ref{fig-lifetime}, SMECN performs consistently
better than MECN for both.  The number of
nodes still alive and the number of nodes still connected to the sink
decrease much more slowly in SMECN than in MECN. For example, in Figure
\ref{fig-lifetime}(a), at time $1200$, $64\%$ of the nodes have died
for MECN while only $22\%$ of the nodes have died for SMECN.
\input{epsf}

Finally, we collected data on average energy consumption per node at the end
of the simulation, on throughput, and on end-to-end delay.
MECN uses 63.4\% more energy per node than SMECN. 
SMECN delivers more than 127\% more packets
than MECN by the end of the simulation, MECN's delivered packets have an
average end-to-end delay that is 21\% higher than SMECN.  
Overall, it is clear that the performance of SMECN is significantly
better than MECN.  We did not simulate the performance of the
network in the absence of an algorithm designed to conserve power
(This is partly because it was not clear what power to choose for
broadcasts. If 
the maximum power is used, performance will be much worse. If
less power is used, the network may get disconnected.)
However, these results clearly show the advantages of using
an algorithm that increases energy efficiency.

\section{Conclusion}
\label{sec-conclusion}
\commentout{
necessary and sufficient conditions to achieve a
minimum-energy communication network. We propose an improvement of the
protocol described by Rodoplu and Meng
\cite{Rodoplu99} that maintains a smaller set of direct transmission
links than MECN 
while still ensuring that there is a 
 minimum-energy path between any source-destination
pair. 
Thus, our protocol is able to lower link maintenance
costs and achieve more power savings. Our protocol is also
computationally simpler.
We prove the correctness of our protocol. In addition, we propose
a simple energy-efficient reconfiguration protocol that maintains the
above minimum energy path property as the network topology changes
dynamically. We demonstrate the performance improvements of our
protocol through simulation.
}
We have proposed a protocol SMECN that computes a network with
minimum-energy than that computed by the protocol MECN of
\cite{Rodoplu99}.  We have shown by simulation that SMECN performs
significantly better than MECN, while being computationally simpler.
 
As we showed in Proposition~\ref{N2char}, 
in the case of a circular search space,
SMECN  computes the set $E_2$ consisting of all edges that are 
not 2-redundant.  
In general, we can find a communication network with
the minimum-energy property that has fewer edges by discarding edges that
are $k$-redundant for $k > 2$.  Unfortunately, for $u$ to compute whether an
edge is $k$-redundant for $k > 2$ will, in general, require information
about the location of nodes that are beyond $u$'s broadcast range.
Thus, this computation will require more broadcasts and longer messages on
the part of the nodes.  There is a tradeoff here; it is not clear that
the gain in having fewer edges in the communication graph is compensated
for by the extra overhead involved. We plan to explore this issue
experimentally in future work.


\section{Acknowledgments}
We thank Volkan Rodoplu at Stanford University for his kind
explanation of his MECN algorithm, his encouragement to publish
these results, and for making his code publicly
available. We thank 
Victor Bahl, Yi-Min Wang, and Roger
Wattenhofer at Microsoft Research for 
helpful discussion.

\bibliographystyle{plain}
\bibliography{lilisbib}
\end{document}